\documentclass[12pt]{article}

\usepackage[a4paper, margin=2.5cm]{geometry}
\usepackage{moreverb}
\usepackage{booktabs}

\usepackage[latin1]{inputenc}
\usepackage{subfig,bm,todonotes, amsmath}

\usepackage{multirow}

\usepackage{graphicx}
\usepackage{lipsum}
\usepackage{amsthm, amsmath, amssymb, amsfonts,verbatim}

\usepackage{graphicx,epstopdf}
\usepackage{algpseudocode}
\usepackage{multirow}
\usepackage{booktabs}
\usepackage{animate}
\usepackage{graphicx}
\usepackage{booktabs}
\usepackage{multirow}
\usepackage{longtable}
\usepackage{enumerate}
\usepackage{color,shortvrb}
\usetikzlibrary{calc,arrows,positioning,matrix,fit,intersections,shapes,shapes.misc,snakes}
\usepackage{float}

\usepackage{lscape}
\usepackage{url}
\usepackage[breaklinks]{hyperref}
\usepackage{breakurl}
\usepackage{authblk}

\begin{document}

\title{\textbf{Bayesian bivariate meta-analysis of diagnostic test studies with interpretable priors}}

\author{J. Guo} 
\author{A. Riebler} 
\author{H. Rue} 
\affil{Department of Mathematical Sciences, Norwegian University of Science and Technology, Trondheim, PO 7491, Norway.}


\maketitle

\begin{abstract}
In a bivariate meta-analysis the number of diagnostic studies involved
is often very low so that frequentist methods may result in problems. 
Bayesian inference is attractive as informative priors that add small
amount of information can stabilise the analysis without overwhelming
the data. However, Bayesian analysis is often 
computationally demanding and the selection of the prior for the covariance matrix of the
bivariate structure is crucial with little data. 
The integrated nested Laplace approximations (INLA) method provides
an efficient solution to the computational issues by avoiding any sampling,
but the important question of priors remain. We explore the penalised complexity (PC) prior 
framework for specifying informative priors for the variance parameters
and the correlation parameter. PC priors facilitate
model interpretation and hyperparameter specification as expert
knowledge can be incorporated intuitively.  We conduct a simulation
study to compare the properties and behaviour of differently defined PC priors 
to currently used priors in the field. The simulation
study shows that the use of PC priors results in more precise
estimates when specified in a sensible
neighbourhood around the truth. 
To investigate the usage of PC priors in practice we reanalyse a meta-analysis using the telomerase marker for the diagnosis of bladder cancer.
\end{abstract}

\section{Introduction}
A meta-analysis summarises the results from several independently
published studies. It plays a central role in scientific areas, such as
medicine, pharmacology, epidemiology, education, psychology,
criminology and ecology \cite{borenstein2011introduction}. Within
epidemiology the application to diagnostic test studies is very common
\cite{reitsma2005bivariate, chu2006bivariate}. Results of a diagnostic
test study typically are presented in a two-by-two table, i.e.~a
cross-tabulation, with four entries: the number of patients tested
positive that are diseased (according to a gold standard), those
tested positive that are not diseased, those tested negative that are
however diseased and finally those tested negative that are not
diseased. Usually the table entries are referred to as true positives
(TP), false positives (FP), false negatives (FN) and true negatives
(TN), respectively. Based on those study-specific entries sensitivity and specificity estimates can be derived. Sensitivity measures the probability of correctly
diagnosing the presence of a disease, while specificity measures the
probability of correctly diagnosing the absence of a disease. The main
goal of a bivariate meta-analysis is to derive summary estimates of
sensitivity and specificity over all studies. Pairs of sensitivity and
specificity are jointly analysed and the correlation between them is 
incorporated using a bivariate random effects approach \cite{reitsma2005bivariate,chu2006bivariate}.

Bivariate meta-analysis is an active area of methodological research
\cite{chu2006bivariate, verde2010meta, paul2010bayesian, 
  menke2013bivariate, eusebi2014latent, zapf2015nonparametric}. One reason is that many diagnostic
meta-analyses combine results from only a small number of studies and
the data involved in a study may be sparse. This makes frequentist
methods challenging since maximum likelihood estimation may have
problems \cite{paul2010bayesian, riley2007bivariate}. Bayesian hierarchical models became attractive because
the use of priors can stabilise the analysis whereby the data will
still dominate if enough information is available. In this way, 
researchers can incorporate the expected degree of between-study
heterogeneity and information based on
comparable studies
\cite{turner2015predictive}. Markov Chain Monte Carlo (MCMC) methods
are frequently used to analyse Bayesian hierarchical models
\cite{menke2013bivariate}. However MCMC often is computationally
intense and not practical, such as in a simulation study where a range
of methods should be compared on a large number of datasets. The selection of the prior for the
covariance matrix of the bivariate structure adds another difficulty,
as the obtained estimates are likely to be sensitive when the data
set is small. In practice, parameters of a prior are often taken
from previous analyses, whereby their suitability for the study of
interest is not always obvious. 

Paul et al. \cite{paul2010bayesian} proposed to use integrated nested Laplace
approximations (INLA) to perform full Bayesian inference for bivariate
meta-analysis \cite{rue2009approximate}. The new R-package \texttt{meta4diag} provides an easier interface to this functionality in the R-package \texttt{INLA} \cite{guo2015}.
Harbord \cite{harbord2011commentary} 
noted that INLA has considerable promise for use in these kind of
models, but specifying suitable prior distributions remains a
difficulty. Traditionally, non-informative or weakly informative
priors have been used. Lambert et al. \cite{lambert2005vague} compared
the performance of 13 different vague priors for the variance
parameters using simulated data from bivariate meta-analysis and
concluded that the results are very sensitive across different priors
when the sample size is small. Thorlund et
al. \cite{thorlund2013modelling} showed that moderately informative
priors have better performance compared to non-informative or weakly
informative priors for estimating the variance parameters. The influence of different priors on the
estimation of the correlation parameter is less clear. Paul et al. \cite{paul2010bayesian}
used a normal prior on the Fisher's z-transformed correlation
parameter, and concluded that a U-shaped prior performs better than an 
approximate uniform prior on the correlation scale. Most priors that are
used for the correlation parameter are centred at a reference value equal to zero and the density
mass around zero is symmetrically distributed. However, since
sensitivity and the specificity in a diagnostic test of meta-analysis
study are commonly assumed to be negatively correlated
\cite{reitsma2005bivariate}, Harbord \cite{harbord2011commentary}
suggested to use a stronger prior for the correlation parameter that
is not symmetric around zero, but defined around a (negative) reference value.

Recently, Simpson et al. \cite{2014arXiv1403.4630S} proposed the new
framework of penalised complexity (PC) priors, which allows such a specification intuitively. 
Here, the bivariate model component is seen as a flexible extension of a base model, i.e. a simpler model, to which it will shrink to if not otherwise indicated by the data. 
Unlike the prior by Paul et al. \cite{paul2010bayesian}, the amount of shrinkage and the reference value in this prior can be
 controlled in an interpretable way using expert knowledge.

In this paper, we use the PC prior framework to motivate 
prior distributions for the variances and the correlation parameter, and compare their performance to commonly used priors through an extensive simulation
study. Section 2 introduces the bivariate meta-analysis model as used
in the paper. Section 3 discusses PC priors for the correlation
parameters and proposes different ways to incorporate prior
knowledge. Further, PC priors for the variance parameters are
presented. Section 4 presents a simulation study to assess the performance of the PC prior. A meta-analysis investigating the use of a telomerase marker for bladder cancer is presented in Section 5. We end the paper with a discussion and concluding remarks in Section 6.

\section{The Bayesian Bivariate Model} 
Chu and Cole \cite{chu2006bivariate} proposed a hierarchical model which
analyses the two-by-two tables of several diagnostic test studies
together. The model has two levels. At the data level  the number of
true positives and the number of true negatives are assumed to be
separately binomial distributed. These two models are linked at the latent
level trough a bivariate random effect which accounts for
between-study heterogeneity.

Let $i=1, \dots, I$ be the study index, and TP, FP, TN and FN denote the number of true positives, false
positives, true negatives, and false negatives, respectively. Further,
let Se and Sp denote sensitivity and specificity, respectively. Then the model is as follows
\begin{equation}
\begin{aligned}
&\text{TP}_i | \text{Se}_i\sim \text{Binomial}(\text{TP}_i+\text{FN}_i,\text{Se}_i),\quad \text{logit}(\text{Se}_i)=\mu + \mathbf{U}_i\mathbf{\alpha}+\phi_i,\\
&\text{TN}_i | \text{Sp}_i\sim \text{Binomial}(\text{TN}_i+\text{FP}_i,\text{Sp}_i),\quad \text{logit}(\text{Sp}_i)=\nu + \mathbf{V}_i\mathbf{\beta}+\psi_i,\\
&\begin{pmatrix} \phi_i \\ \psi_i \end{pmatrix} \sim \mathcal{N}\left[ \begin{pmatrix} 0 \\ 0 \end{pmatrix},\begin{pmatrix} \sigma_{\phi}^2 & \rho\sigma_{\phi}\sigma_{\psi} \\ \rho\sigma_{\phi}\sigma_{\psi} & \sigma_{\psi}^2 \end{pmatrix}  \right],
\end{aligned}
\label{eq1}
\end{equation}

\noindent where  $\mu$, $\nu$ are intercepts for $\text{logit}(\text{Se}_i)$ and $\text{logit}(\text{Sp}_i)$, respectively, and $\mathbf{U}_i$, $\mathbf{V}_i$ are covariate vectors, if available, and $\alpha$ and $\beta$ denote the corresponding regression coefficents. The covariance matrix of the random effects parameters $\phi_i$ and $\psi_i$ is parameterised using between-study
variances $\sigma_{\phi}^2$, $\sigma_{\psi}^2$ and between-study correlation $\rho$.

Another level is added to specify hyperpriors for
all parameters. For the intercepts $\mu$ and $\nu$, as well as for
potential regression parameters it is common to use normal priors with
mean zero and variance fixed to a large value, e.g.~100. Specifying
prior distributions for the between-study variances $\sigma_{\phi}^2$
and $\sigma_{\psi}^2$ and the between-study correlation $\rho$ is
challenging. In some applications  an inverse Wishart distribution is
used for the entire covariance matrix due to its conjugacy for the
normal model \cite{verde2010meta, menke2013bivariate,
  alvarez2014bayesian}. However, using an inverse Wishart prior the correlation estimates might
depend on the value of the variances, which may lead to overestimated correlation estimates. See \cite{alvarez2014bayesian, bartheleme-blogg} for a discussion.

Here, we follow Wei and Higgins \cite{wei2013bayesian} and specify
priors for the variance components and the correlation component
separately. Barnard et al. \cite{barnard2000modeling} proposed a
separation strategy to separate the components in a covariance matrix as
\begin{equation}
\mathbf{\Sigma} = \begin{pmatrix} \sigma_{\phi}^2 & \rho\sigma_{\phi}\sigma_{\psi} \\ \rho\sigma_{\phi}\sigma_{\psi} & \sigma_{\psi}^2 \end{pmatrix} = \mathbf{V}^{1/2}\mathbf{R}\mathbf{V}^{1/2} = 
\begin{pmatrix} \sigma_{\phi} & 0 \\ 0 & \sigma_{\psi} \end{pmatrix}\begin{pmatrix} 1 & \rho \\ \rho & 1 \end{pmatrix}\begin{pmatrix} \sigma_{\phi} & 0 \\ 0 & \sigma_{\psi} \end{pmatrix},
\label{eq:barnard}
\end{equation}

where $\mathbf{V}^{1/2}$ is a diagonal matrix with the standard
deviations as elements and $\mathbf{R}$ is the $2\times 2$ matrix of
correlations. With this separation we can define priors for the
variances and the correlation separately. Mainly vague or mildly
informative priors for $\sigma^2_{\phi}$, $\sigma^2_{\psi}$ and $\rho$
are used \cite{verde2010meta, paul2010bayesian,
  jackson2011multivariate}. For the variance parameters
$\sigma^2_{\phi}$ and $\sigma^2_{\psi}$ inverse gamma priors are
commonly chosen, whereby the value of the hyperparameters are often
copied from the literature.  Thorlund et
al. \cite{thorlund2013modelling} found that the use of mildly
informative priors for the variance components of a bivariate
meta-analysis yielded a better model fit and narrower credible
intervals compared to weakly informative and non-informative
priors for the correlation parameter. Paul et al. \cite{paul2010bayesian} reached a similar
conclusion, when they compared three differently shaped priors. They
proposed to use an U-shaped prior for the correlation parameters
instead of a roughly uniform prior to obtain unbiased estimates. It is common to use a normal
distribution on the Fisher's z-transformed correlation
$\theta=\text{logit}((\rho+1)/2)$ that is centered at
0 as prior for the correlation parameter. However, in a bivariate meta-analysis of diagnostic test studies we
expect sensitivity and specificity to be negatively correlated
\cite{reitsma2005bivariate}. For this reason, Harbord
\cite{harbord2011commentary} proposed to use a stronger prior for
$\rho$ that is not symmetric around zero, but defined around a
(negative) constant $\rho_0$. Incorporating this prior belief may
additionally stabilise the analysis if few studies are
available. Here, we use the framework of penalised complexity (PC)
priors that will allow us to address Harbords suggestion in an
interpretable way. 
%


\section{Penalised Complexity Priors} 
\label{sec:pcprior}
The construction of PC priors is based on four principles: 1. Occam's razor -  a simpler model, called base model, should be preferred until there is enough information to support a more complex model. 2. Measure complexity -  the Kullback-Leibler divergence is used to measure model complexity. 3. Constant-rate penalisation - the deviation from the base model is penalised using a constant decay rate. 4. User-defined scaling - the user has some idea of the range in which the parameter of interest lies. 

\subsection{PC priors for the correlation parameter}
Often a normal distribution is placed on the Fisher's z-transformed correlation, i.e., $\theta\sim\mathcal{N}(\mu,\sigma^2)$, and suitable values for $\mu$ and $\sigma^2$ have to be chosen. Paul et al. proposed to use $\mu=0$ and $\sigma^2=5$. The prior mean $\mu$ could be easily changed to a negative value, say, however the density distribution weight on both sides of this value will still be the same. It seems more natural to define a prior that satisfies $P(\rho \leqslant \rho_0)=\omega$, where $\rho_0$ is a (negative) value and $\omega$ denotes the prior probability that $\rho$ is smaller than~$\rho_0$.

\subsubsection{Equal probability density on both sides of $\rho_0$}
Consider the bivariate random effects model in Equation~(\ref{eq1}), where the bivariate random effect is assumed to be normally distributed with zero-mean and covariance matrix $\mathbf{\Sigma}$. We follow the separation strategy given in Equation \eqref{eq:barnard} and consider only the correlation matrix $\mathbf{R}$ in this section. Let $\mathcal{N}(\mathbf{0}, \mathbf{R}_{b})$ and $\mathcal{N}(\mathbf{0}, \mathbf{R}_{f})$ denote the base and flexible model, respectively, where

\begin{align*}
	\mathbf{R}_{b} = \begin{pmatrix}1 & \rho_0 \\ \rho_0 & 1\end{pmatrix}\quad \quad \quad\quad\quad
 \text{and} \quad \quad \quad\quad\quad
	\mathbf{R}_{f} = \begin{pmatrix}1 & \rho \\ \rho & 1\end{pmatrix},
\end{align*}
where $\rho_0$ is fixed to a (negative) value in which one believes with respect to the application, while $\rho$ is a random variable. Thus, $\mathbf{R}_{b} $ is a special case of $\mathbf{R}_{f} $, to which we would like to shrink if not otherwise indicated by the data.
The increased complexity introduced by $\mathcal{N}(\mathbf{0}, \mathbf{R}_{f})$
compared to $\mathcal{N}(\mathbf{0}, \mathbf{R}_{b})$ is measured
by the Kullback-Leibler divergence (KLD). The KLD is a measure of the `information" lost when the base model $\mathcal{N}(\mathbf{0}, \mathbf{R}_{b})$ is used to approximate the more flexible model $\mathcal{N}(\mathbf{0}, \mathbf{R}_{f})$
\begin{align*}
\text{KLD}\left( \mathcal{N}(\mathbf{0}, \mathbf{R}_{f}) || \mathcal{N}(\mathbf{0}, \mathbf{R}_{b})  \right)=\text{KLD}(\rho) &= \frac{1}{2} \left\{ \text{tr}(\mathbf{R}_{b}^{-1}\mathbf{R}_{f}) -2-\ln\left(\frac{|\mathbf{R}_{f}|}{|\mathbf{R}_{b}|} \right) \right\}\\[0.2cm]
&= \frac{1-\rho_0\rho}{1-\rho_0^2} -1 -\frac{1}{2} \ln (1-\rho^2) + \frac{1}{2} \ln (1-\rho_0^2),
\end{align*}
and $d(\rho)=\sqrt{ 2\text{KLD}(\rho) }$ is defined as the ``distance" between the base model and the flexible model \cite{2014arXiv1403.4630S}. A distance equal to zero means the flexible model shrinks to the base model, whereas increasing distance indicates that the flexible model moves further away from the base model. The base model is preferred unless there is enough supporting evidence from the data. Therefore a prior distribution on the distance scale should have mode at $0$ and the probability density should decrease with distance. Any distribution that fullfills those features can be used as prior distribution on the distance. Simpson et al. \cite{2014arXiv1403.4630S} proposed a constant rate penalisation for the distance $d$. This leads to an exponential prior with parameter $\lambda$ which controls the decay-rate of the deviation from the base model. Using a change of variables we derive the prior for the correlation as $\pi(\rho)=\lambda\exp(-\lambda d(\rho))|\partial d(\rho)/\partial \rho|$. It is worth noting that while the distance goes from zero to infinity, we can deviate from the base model in two directions, either from $\rho_0$ to $1$ or from $\rho_0$ to $-1$. The exponential prior thus is a two-sided prior that shares the same parameter $\lambda$ and has identical behaviour on the distance scale for $\rho\in [-1, \rho_0]$ and $\rho\in [\rho_0,1]$, leading to the same probability weights on both sides of $\rho_0$ in the correlation scale. 

\subsubsection{Unequal probability density on both sides of $\rho_0$}
In order to distribute the density unevenly to each side of $\rho_0$ but keeping the assumption of a constant rate penalisation, we use two rate parameters $\lambda_1$ and $\lambda_2$ which indicate the decay-rate of the deviation from the base model when $\rho\leqslant \rho_0$ and $\rho >\rho_0$, respectively. In order to merge these two exponential priors  $\pi_1(d;\lambda_1)$ and $\pi_2(d;\lambda_2)$ in the distance scale into one continuous prior distribution $\pi(\rho; \lambda_1,\lambda_2)$ in the correlation scale, we define
\begin{equation*}
\pi(\rho;\lambda_1,\lambda_2) = \left\{
\begin{aligned}
\omega_1\pi_1(d(\rho);\lambda_1)\left| \frac{\partial d(\rho)}{\partial \rho} \right|,\quad \quad \rho\leqslant \rho_0\\
\omega_2\pi_2(d(\rho);\lambda_2)\left| \frac{\partial d(\rho)}{\partial \rho} \right|,\quad \quad \rho> \rho_0
\end{aligned}
\right.
\end{equation*}
where $\omega_1=\int_{-1}^{\rho_0}\pi(\rho;\lambda_1,\lambda_2)$ and $\omega_2 = \int_{\rho_0}^{1}\pi(\rho;\lambda_1,\lambda_2)\in [0, 1]$, are the probability of $\rho\leqslant \rho_0$ and $\rho >\rho_0$, respectively. To obtain a proper probability density, $\omega_1$ and $\omega_2$ must satisfy $\omega_1+\omega_2=1$. Further, to get a continuous probability density for $d=0$, $\omega_1$ and $\omega_2$ must satisfy $\omega_1\lambda_1 = \omega_2\lambda_2$.

The value of the hyperparameters $\lambda_1$ and $\lambda_2$ can be specified by defining the density behaviour around $\rho_0$.  Consider for example the two conditions $P(\rho\leqslant u_{\text{min}}|-1<u_{\text{min}}<\rho_0)=\alpha_1$, where $u_{\text{min}}$ is a ``left tail event" and $\alpha_1$ is a small probability of this event, and $P(\rho\leqslant\rho_0)=\omega_1$ which is equivalent to $P(\rho>\rho_0)=\omega_2$. Solving $\int_{u_{\text{min}}}^{\rho_0}\pi(\rho;\lambda_1,\lambda_2)\text{d}\rho=\omega_1-\alpha_1$ and $\omega_1\lambda_1 = (1-\omega_1)\lambda_2$ we find
\begin{align*}
\lambda_1 = \frac{\log(\omega_1)-\log(\alpha_1)}{d(u_{\text{min}})}\quad \quad\quad\quad\quad \text{and} \quad \quad\quad\quad\quad \lambda_2 = \frac{\omega_1\lambda_1}{1-\omega_1},
\end{align*}
see Appendix A. Instead of the two conditions presented, alternative conditions could be  specified. Here we provide three different strategies to choose $\lambda_1$ and $\lambda_2$ based on prior knowledge, see Table~\ref{table:strategies}. For all strategies we first need to set $\rho_0$ which represents a reference value. A priori we believe that the true correlation $\rho$ is likely to be close to $\rho_0$. For most bivariate meta-analyses $\rho_0$ would be set negative, however it could also be set equal to zero or even positive if sensible for the application considered. The details to find $\lambda_1$ and $\lambda_2$ under the different strategies are given in Appendix A.

\subsubsection{Choice of hyperparameters}
\begin{figure}[h!]
\begin{minipage}[t]{\linewidth} 
\centering
\includegraphics[width=0.66\textwidth]{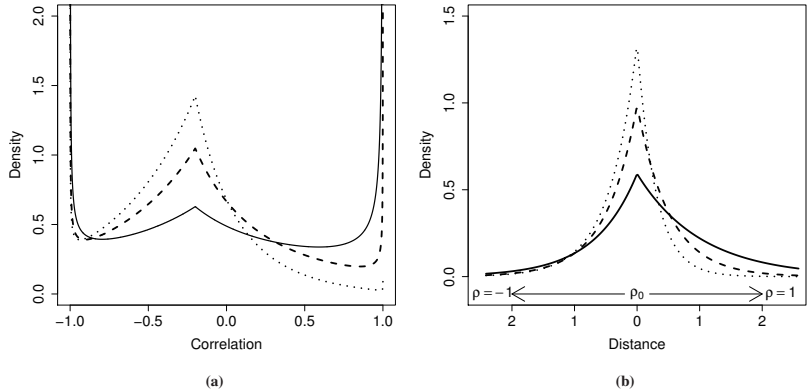}
\caption{Shown are the PC 1, PC 2 and PC 3 prior in solid, dashed and dotted lines, respectively. Panel (a) shows the prior distributions in the correlation scale, Panel (b) shows the prior distributions in the distance scale. All PC priors have reference value at $\rho_0=-0.2$. }
\label{guo1}
\end{minipage}
\end{figure}

The choice of the parameters for the different strategies is still a delicate issue, in particular if the meta-analysis consists of few studies and the number of patients is low. Ideally, there is expert knowledge available to set up sensible probability contrasts. Depending on how concrete the expert knowledge is, the informativeness of the prior distribution can be controlled. Alternatively, we could motivate the probability contrasts based on  a collection of existing meta-analyses. Menke \cite{menke2010bivariate}  studied 50 independent bivariate meta-analyses which were selected randomly from the literature within a Bayesian setting, whereas Diaz \cite{diaz2015performance} reported frequentist estimates based on a literature review of 61 bivariate meta-analyses of diagnostic accuracy published in 2010. Based on these two publications, the distribution of the correlation seems asymmetric around zero. We find that around half of the correlation point estimates are negative, with a mode around $-0.2$. Only a small proportion are larger than $0.4$ and values larger than $0.8$ are rare.  Based on these findings, we choose three differently behaved PC priors that are defined around $\rho_0=-0.2$:
\begin{center}
\begin{tabular}{p{2cm}p{11cm}}
PC 1:  & $\rho\sim PC(\rho_0=-0.2,\  \omega_1=0.4,\  u_{\text{min}}=-0.95,\  \alpha_1=0.05)$  \\[0.2cm]
PC 2:  & $\rho\sim PC(\rho_0=-0.2,\  u_{\text{min}}=-0.9,\  \alpha_1=0.05,\  u_{\text{max}}=0.8,\  \alpha_2=0.05)$  \\[0.2cm]
PC 3:  & $\rho\sim PC(\rho_0=-0.2,\  \omega_1=0.6,\  u_{\text{max}}=0.4,\  \alpha_2=0.05)$  \end{tabular}
\end{center}
Figure~\ref{guo1} shows all priors in the original scale and the distance scale. PC 1 is the least informative prior, while PC 3 is the most informative prior with almost no support for correlation values close to one (panel a)). We see that all priors go to infinity at the boundary values -1 and 1. This is a consequence of the bounded parameterisation. Looking at the the distance scale we see a mode at the base model, i.e. at distance equal to zero. The density support in the distance scale is positive, however we can deviate from $\rho_0$ in two different directions at possibly different decay rates. The right side of panel b) shows the prior when deviating towards $\rho=1$ with rate $\lambda_2$, while the left side shows the resulting prior when deviating towards $\rho=-1$ with rate $\lambda_1$. The density decreases towards zero the further we deviate from the base model, the larger the distance to $\rho_0$ gets.

\begin{figure}[h!]
\begin{minipage}[t]{\linewidth} 
\centering
\includegraphics[width=0.99\textwidth]{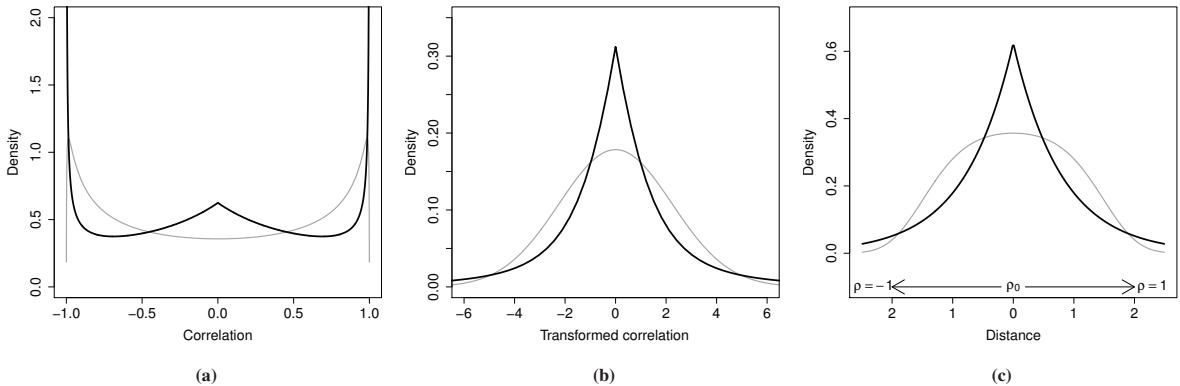}
\caption{Shown are the Paul et al. prior and PC 0 prior for the correlation parameter in gray and black lines, respectively. Panel (a) shows the prior distributions in the correlation scale, Panel (b) shows the prior distributions in the Fisher's z-transformed correlation scale and Panel (c) in the distance scale. }
\label{guo2}
\end{minipage}
\end{figure}

We compare their performance to the Paul et al. prior and a less informative PC-prior which, similar to the Paul et al. prior, also fulfils $P(\rho<0)=0.5$ and $P(\rho< -0.9)=0.1$:
\begin{center}
\begin{tabular}{p{2cm}p{11cm}}
PC 0:  & $\rho\sim PC(\rho_0=0,\  \omega_1=0.5,\  u_{\text{min}}=-0.9,\  \alpha_1=0.1)$  \\[0.2cm]
Paul et al.: & $\theta \sim \mathcal{N}(\mu = 0, \  \sigma^2=5)$  \\[0.2cm]
\end{tabular}
\end{center}
Figure~\ref{guo2} shows both priors. In addition to the original and the distance scale, we also present the priors on the Fisher's z-transformed correlation scale on which the Paul et al. prior is defined. The PC 0 prior is less informative compared to the priors shown in Figure~\ref{guo1}, however compared to the Paul et al. prior it is more informative which becomes apparent in all three scales by a more concentrated shape around the base model. The flatness of the Paul et al. prior in the distance scale implies that in an area around the base model it behaves almost like a uniform prior, which means the base model is not necessarily preferred. In the correlation scale we see that the PC 0 prior goes to infinity at the boundaries whereas the Paul et al. prior goes to zero.  This is the result of a different behaviour when doing the change from variables from an infinite to a finite support. 

\subsection{PC priors for the variance parameters}
Here, we illustrate the PC priors for the variance components $\sigma_{\phi}^2$ or $\sigma_{\psi}^2$ in model (\ref{eq1}). In Section 3.2  of \cite{2014arXiv1403.4630S}, Simpson et al. derived the PC-prior for a precision parameter which is the inverse of the variance parameter. It was found to be type-2 Gumbel distribution which leads to a Weibull distribution with shape parameter equal to $1/2$ for the variance parameter. The second parameter depends on the rate of the decay $\lambda$. A simple choice to set $\lambda$ is to provide $(u, a)$ such that $P(\sigma>u)=a$ leading to an explicit relationship between distribution parameter and density behaviour $\lambda=-\log(u)/a$. The magnitude of the penalty for deviating from the base model $\sigma^2\rightarrow0$ can be controlled through setting the value of the parameter $\lambda$, and it is thus controlled by $u$ and $a$.

\begin{figure}[h!]
\begin{minipage}[t]{\linewidth} 
\centering
\includegraphics[width=0.66\textwidth]{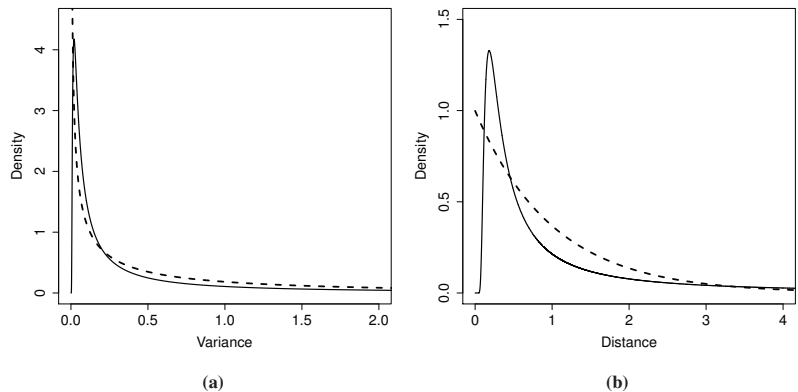}
\caption{Illustration of the PC prior for a variance parameter calibrated such that $P(\sigma>3)=0.05$ (dashed) compared to an $\Gamma^{-1}(\text{shape} = 0.25, \text{rate} = 0.025)$ (solid). Panel (a) shows both densities in variance scale while Panel (b) shows  them on the distance scale.}
\label{guo3}
\end{minipage}
\end{figure}

Figure~\ref{guo3} shows an example of a PC prior for a variance parameter defined under the condition $P(\sigma>3)=0.05$. This choice corresponds to the belief that the sensitivities or specificities lie in the interval $[0.5, 0.95]$ with probability $0.95$. We compare this prior with an inverse gamma prior with shape parameter equal to $0.25$ and rate parameter equal to $0.025$, i.e., $\Gamma^{-1}(0.25, 0.025)$, as proposed by Paul et al. \cite{paul2010bayesian}. In the distance scale it can be seen that the inverse gamma prior has zero density at distance zero, which means that it is not possible to shrink to the base model $\sigma^2 \rightarrow 0$. Hence the inverse gamma prior supports overfitting in the sense that a too complex model is favoured even if not supported by the data.

\section{Simulation Studies}
In this section, we conduct an extensive simulation study to compare the effects on the posterior marginals when using different priors. For this purpose we simulate from model~(\ref{eq1}) as described in \cite{paul2010bayesian}. We use the following specifications:
\begin{enumerate}
\item The number of individuals per study $n_i$ (composed of diseased and non-diseased individuals) are sampled from a shifted gamma distribution and round off to the nearest integer;
\item The number of studies are 10, 25, 50;
\item The true mean values of sensitivity and specificity are $80\% / 70\%$, $90\% / 90\%$, and $95\% / 30\%$;
\item Correlation between logit transformed sensitivity and specificity across studies: -0.95, -0.8, -0.6, -0.4, -0.2, 0, 0.2, 0.4, 0.6;
\item The true variances of logit transformed sensitivity and specificity are $1$ for all the scenarios.
\end{enumerate}
This leads to a total of 81 different scenarios. The detailed specifications are shown in Table~\ref{table:table1}. For each scenario 1000 data sets are generated and for each data set summary estimates of all posterior marginals using different priors are computed. We compare the results using  errors (the difference between the posterior median and the true value), bias (mean errors), MSE (sum of squared bias and squared standard deviation of the posterior medians), and coverage probabilities (frequency in which the true value is within the $95\%$ credible interval).

\subsection{Simulation Result}
\subsubsection{Comparison of different priors for the variance parameters}
\begin{figure}[h!]
\begin{minipage}[t]{\linewidth} 
\centering
\includegraphics[width=0.99\textwidth]{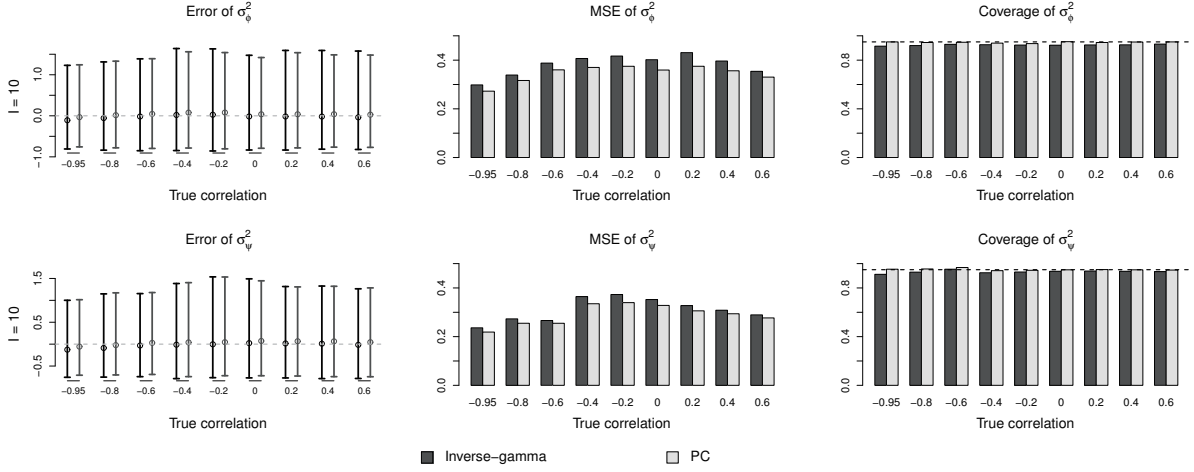}
\caption{Errors, bias, MSE and $95\%$ coverage probabilities for $\sigma^2_{\phi}$ (top row) and $\sigma^2_{\psi}$ (bottom row). The dark gray bars in each plot  indicate the results using $\Gamma^{-1}(0.25,0.025)$ prior whereas the light gray bars are for $PC(u=3,\alpha=0.05)$. The dashed line in the coverage plot corresponds to the nominal level of $95\%$. The prior for the correlation parameter is always $\theta\sim\mathcal{N}(0,5)$. The results are obtained using INLA from the data simulated with $I=10$, $\mu=0.8$, $\nu=0.7$, $\sigma^2_{\phi}=\sigma^2_{\phi}=1$ and $\rho=(-0.95,-0.8,-0.6,-0.4,-0.2,0,0.2,0.4,0.6)$.}
\label{guo4}
\end{minipage}
\end{figure}

To study the effect of changing the prior for the variance parameters, the prior for correlation parameter is fixed to the Paul et al. prior. We compare the result from a PC prior with $u=3$ and $\alpha=0.05$ with an inverse gamma prior $\Gamma^{-1}(0.25,0.025)$. We find that changing the priors for the variance parameters does not affect the posterior marginal of the correlation parameter. Figure~\ref{guo4} shows errors, MSE and $95\%$ coverage probabilities for both settings when using $I=10$ studies. From the error plot, we can see that there is little difference between these two priors and the bias, indicated as circle, is almost the same. However, we observe a smaller MSE and marginally better coverage probabilites when using the PC prior for the variance parameters. Increasing the number of studies we observer similar results but the differences decrease slightly, see Supplementary Material. Changing the simulated values of sensitivity and specificity has no influence on the results. In the following we will use the PC prior with $u=3$ and $\alpha=0.05$ for both variance parameters.

\subsubsection{Comparison of different priors for the correlation parameter}
\begin{figure}[h!]
\begin{minipage}[t]{\linewidth} 
\centering
\includegraphics[width=0.99\textwidth]{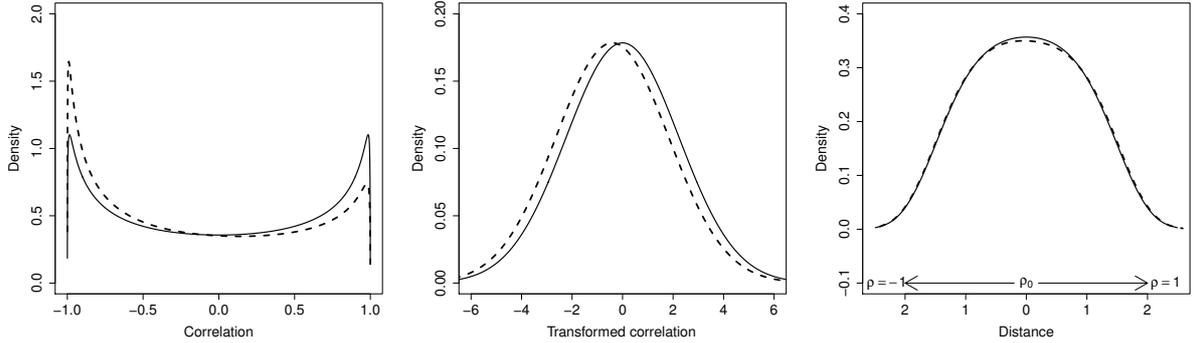}
\end{minipage}
\caption{Shown are the Paul et al. prior (solid) and the shifted normal prior (dashed) for the Fisher's z-transformed correlation parameter $\theta$ in the correlation, transformed correlation and distance scale.}
\label{guo5}
\end{figure}

Next we study the effect of changing the prior for the correlation parameter. Arguing that a central innovation is to move the reference value from zero to a negative value we may just centre the normal distribution for $\theta$ around a negative value, e.g. $\text{logit}(-0.2)$, instead of zero. Figure~\ref{guo5} shows the normal prior proposed by Paul et al. and the shifted normal prior both with variance equal to $5$ in the correlation, transformed correlation and distance scale. There is almost no difference in the distance scale. However in the correlation scale the shifted normal distribution leads to more density mass in the left tail and less in the right tail, while keeping the density in the central region approximately uniform distributed. For this reason, posterior marginals stay almost unchanged using these two priors. In the following, we compare the Paul et al. prior with differently chosen PC priors. 

\begin{figure}[h!]
\begin{minipage}[t]{\linewidth} 
\centering
\includegraphics[width=0.99\textwidth]{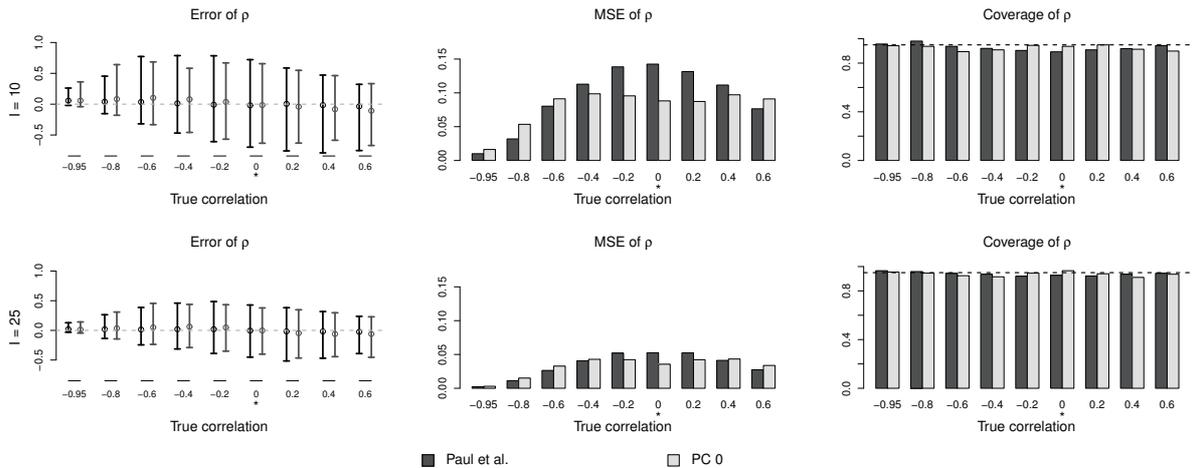}
\caption{Errors, bias, MSE and $95\%$ coverage probabilities for the correlation parameter. The dark gray bars in each plot indicate the results using the Paul et al. prior whereas the light gray bars are for PC 0. The dashed line in the coverage plot corresponds to the nominal level of $95\%$. The small star under zero indicates the base model $\rho_0=0$. The prior for the varaince parameter is always $PC(u=3, \alpha=0.05)$. The results are obtained using INLA from the data simulated with $I=10$ (top row), $I=25$ (bottom row) and $\mu=0.8$, $\nu=0.7$, $\sigma^2_{\phi}=\sigma^2_{\phi}=1$ and $\rho=(-0.95,-0.8,-0.6,-0.4,-0.2,0,0.2,0.4,0.6)$.}
\label{guo6}
\end{minipage}
\end{figure}

\noindent \textbf{Less-informative priors: PC prior vs. Paul et al. prior}
We start by comparing PC 0 with the Paul et al. prior (see Figure~\ref{guo2} for an illustration of both priors).
Both priors provide a small amount of information at the boundaries. We found that using PC 0 leads to a smaller MSE compared to the Paul et al. prior for the variance parameters when $I=10$ in all settings, see Supplementary Material. Bias and coverage probabilities are almost the same. When the number of studies increases, differences in the MSE will also disappear. The resulting effects on the correlation parameter are displayed in Figure~\ref{guo6} for simulation scenarios 1-18. The results for  the remaining scenarios are comparable and available from the authors on request. We find that the posterior results are approximately point-symmetric around $0$ due to the symmetry of the prior. The black circles in the left column indicate that using the Paul et al. prior leads to less biased estimates, however the error bars of PC 0 (gray) indicate a smaller spread when the true correlation is close to the assumed base model, i.e. $0$, also with almost zero bias. From the MSE and $95\%$ coverage probability plot, it is obvious that using PC 0 we obtain a smaller MSE and better coverage probabilities when the true correlation is in a sensible range around the base model.  The spread of errors decreases for both priors when the absolute correlation increases. This is because the correlation estimates are then bounded on one side, whereas the estimates of the non-extreme correlation have two sides to vary. The described results apply to both $I=10$ (top row) and $I=25$ (bottom row) but become less extreme as the number of studies increases.

\begin{figure}[h!]
\begin{minipage}[t]{\linewidth} 
\centering
\includegraphics[width=0.99\textwidth]{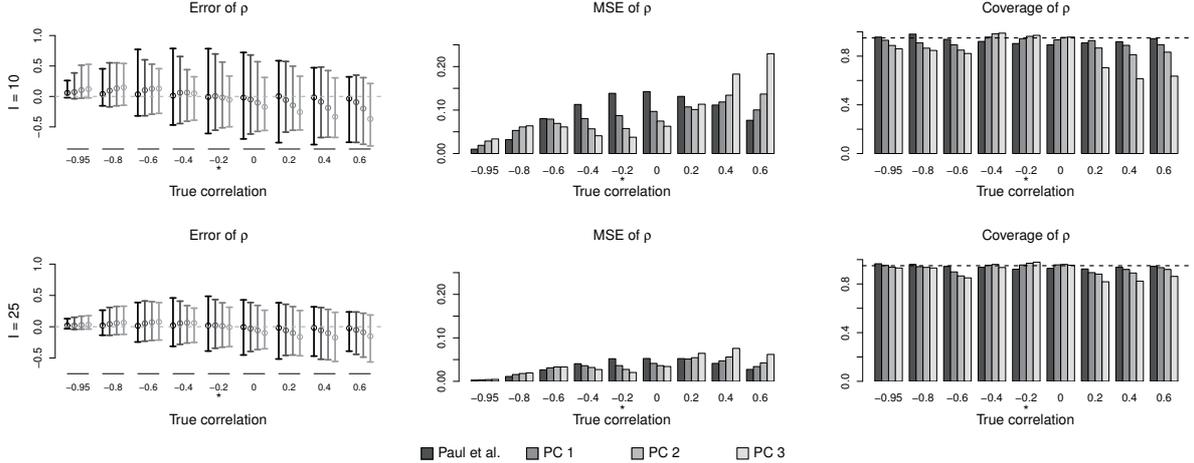}
\end{minipage}
\caption{Errors, bias, MSE and $95\%$ coverage probabilities for the correlation parameter. The dark gray bars in each plot indicate the results using the Paul et al. prior whereas the light gray bars are for PC 1, PC 2 and PC 3. The dashed line in the coverage plot corresponds to the nominal level of $95\%$. The small star under $-0.2$ indicates the base model $\rho_0=-0.2$. The prior for the varaince parameter is always $PC(u=3, \alpha=0.05)$. The results are obtained using INLA from the data simulated with $I=10$ (top row), $I=25$ (bottom row) and $\mu=0.8$, $\nu=0.7$, $\sigma^2_{\phi}=\sigma^2_{\phi}=1$ and $\rho=(-0.95,-0.8,-0.6,-0.4,-0.2,0,0.2,0.4,0.6)$.}
\label{guo7}
\end{figure}

\noindent \textbf{Informative priors vs. less-informative priors} Next we consider PC priors specified at a negative reference value $\rho_0=-0.2$ and having different behaviour on both sides of $\rho_0$. 
Choosing $\rho_0$ means that we change the base model assuming a negative correlation as natural. We compare PC 1, PC 2 and PC 3 as specified in Section~\ref{sec:pcprior} with the Paul et al. prior for the transformed correlation. The estimates of sensitivity and specificity are not sensitive to changes in the prior distributions and stay almost unchanged. If more studies are available, estimates are more precise, the absolute bias is smaller and the MSE is lower for all priors. For the variance parameters, we observe almost the same bias and $95\%$ coverage probabilities for all priors. However, the MSE using PC priors is slightly lower for $I=10$ than when using the Paul et al. prior, see Supplementary Material. Differences disappear with increasing number of studies. Figure~\ref{guo7} displays the results for the correlation parameter for scenarios 1-18. The error estimates obtained by the PC priors are no longer symmetric around the base model as the priors behave differently on both sides of $\rho_0$. From PC 1 to PC 3 priors get more concentrated around the base model, i.e. they are more informative. This results in improved performance measures when the true correlation is close to the base model. If we move further away from the base model the performance in particular the bias decreases. However, the true correlation must be quite far away to obtain worse measures compared to the Paul et al. prior.  In contrast to the PC priors, the Paul et al. prior behaves constantly well in terms of coverage probabilities but performs worse in terms of MSE around the base model.

\section{Diagnosis of bladder cancer using the telomerase marker}
In this section, we apply the methodology to a dataset presented and analysed by Glas et al. \cite{glas2003tumor}. The dataset contains a meta-analysis of 10 studies in which the telomerase marker is used for the diagnosis of bladder cancer, see Table~\ref{table:realdata}

This data set has been analysed in various publications and estimation problems were found when using frequentist approaches \cite{paul2010bayesian, riley2007bivariate}. Riley et al. \cite{riley2007bivariate} used SAS PROC NLMIXED and tried different starting values resulting in an estimated between-study correlation of $-1$ and
 95\% confidence intervals from $-1$ to $1$. The small number of studies and the partly rare data lead to instabilities in the estimation procedure. Using a Bayesian approach prior information can be used to stablize parameter estimation \cite{paul2010bayesian}. Here, we use PC priors to incorporate intuitively prior information into the analysis. For this application we keep the PC(3, 0.05) prior for the variance parameter as motivated in Section 3, as we believe that both sensitivity and specificity lie with $95\%$ probability in the range $[0.5, 0.95]$. Motivating the prior for the correlation is more challenging. Here, knowledge from a clinical epidemiologist could be directly incorporated. We believe that a negative base model is sensible due to potential threshold variations between studies and keep $\rho_0 = -0.2$ \cite{glas2003tumor}. At the same time, we have no strong knowledge regarding the variation around this value and assign the PC 1 prior for the correlation parameter. This prior is more informative than PC 0 but less than PC 2 and PC 3.  The model and all prior settings are specified in the R-package \texttt{meta4diag} which uses INLA for full Bayesian inference \cite{guo2015}. The R-code for implementing the model is shown and explained in Appendix B. Alternatively, the graphical user interface can be used, see Appendix B for a screenshot. 
 
\begin{figure}[h!]
\centering
\includegraphics[width=0.5\textwidth]{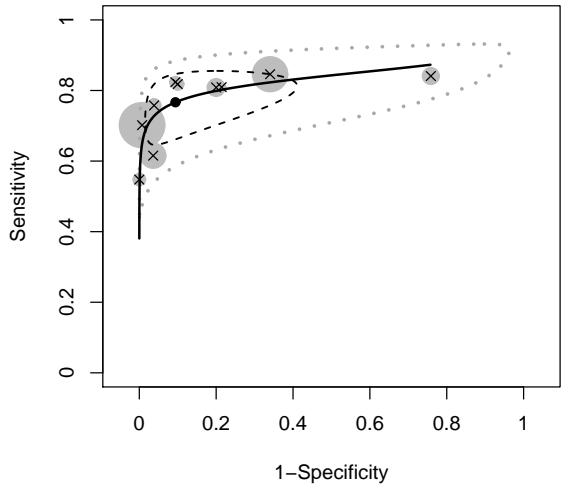}
\caption{SROC plot for telomerase data obtained from meta4diag using the PC 1 prior for the correlation parameter and the PC$(3,0.05)$ prior for the variance parameters. Shown are observed study-specific sensitivity and specificity values (black cross: study sizes are indicated by the gray bubble), SROC line (black solid line), overall sensitivity and specificity estimates from the bivariate model (black point), credible region (black dashed line) and prediction region (gray dotted line).}
\label{guo9}
\end{figure}

Figure~\ref{guo8} shows the posterior marginal distributions for all parameters compared to MCMC samples obtained by JAGS \cite{plummer2003jags}.  The MCMC approach needed about 10 min for 10 000 samples while INLA needed 2 seconds on the same machine. It can be seen that INLA and MCMC coincide very well.
\begin{figure}[h!]
\begin{minipage}[t]{\linewidth} 
\centering
\includegraphics[width=0.99\textwidth]{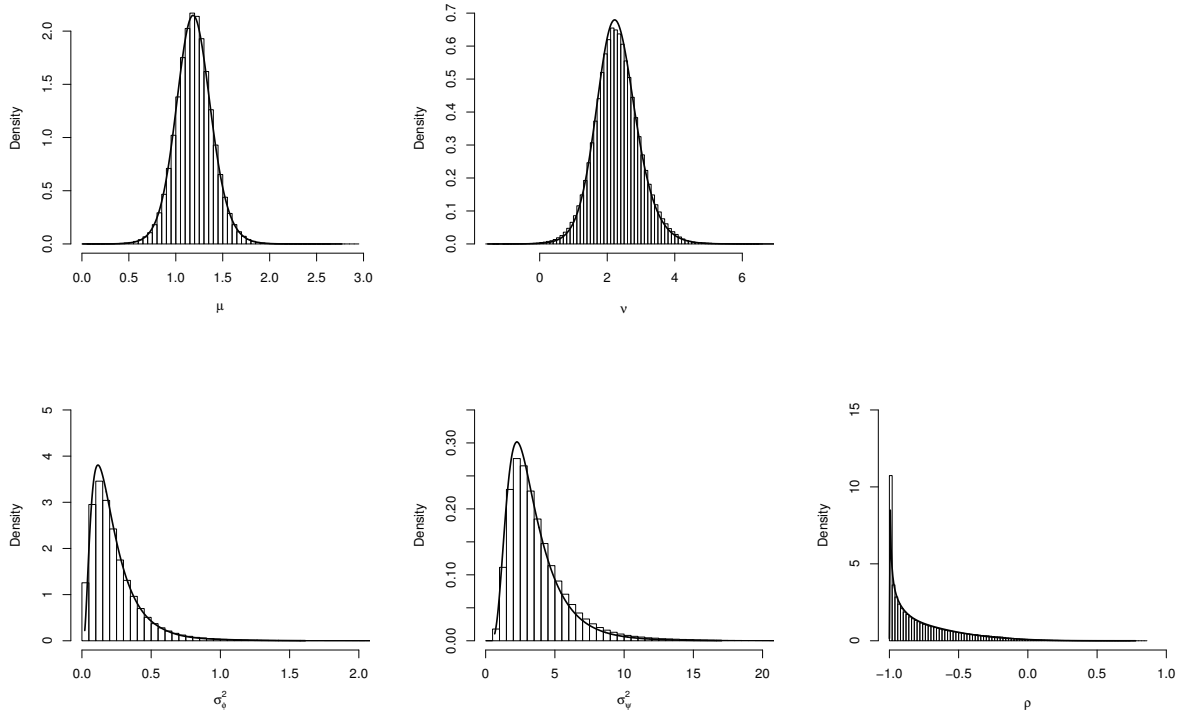}
\end{minipage}
\caption{Posterior marginals obtained by INLA and corresponding histograms of 1 000 000 samples obtained from a MCMC run with 5 000 burn-in
iterations in JAGS. As prior distributions the PC 1 prior for the correlation parameter and the PC prior with hyperparameter $u=3$ and $\alpha=0.05$ for the variance parameters were used.}
\label{guo8}
\end{figure}
The posterior summary estimates for sensitivity and specificity obtained by INLA are 
$0.77$ ($95\%$ CI: [0.71, 0.82])  and $0.91$ ($95\%$ CI: [0.79, 0.97]), respectively, and are shown as summary point (black filled point) in Figure~\ref{guo9}. The posterior correlation was estimated to $-0.86$ ($95\%$ CI: [-0.99, -0.20]) and is slightly lower than $-0.89$ which was obtained by Paul et al.. However, the credible interval is slightly smaller. The strong negative correlation indicates a threshold effect which agrees with the findings of Glas et al. \cite{glas2003tumor}. As there are no covariates involved the parameter estimates can be transformed to the output of an SROC analysis \cite{rutter2001hierarchical}. The resulting SROC curve together with a joint $95\%$ credible and prediction region is shown in Figure~\ref{guo9}.

\section{Discussion}

One of the main challenges of bivariate meta-analysis is that often only few
studies are available and/or data within these studies are
sparse. Frequentist approaches may encounter problems leading to
unreliable parameter estimates such as correlation estimates close to
the boundary. There exist different approaches to address this. For example, Kuss~et~al.~\cite{kuss2014meta} use marginal beta-binomial distributions for the true positives and the true negatives, linked by copula distributions. They found the model to be superior to the bivariate meta-regression model estimated in a frequentist setting. However, the results are sensitive with respect to the copula class used and expert knowledge is hard to incorporate.  Zapf~et~al.~\cite{zapf2015nonparametric} propose a nonparametric frequentist model which shows good convergence properties but does not yet allow the incorporation of covariate or expert knowledge. In this paper we follow an alternative way to stabilise parameter estimation and use a Bayesian approach where additional information is incorporated into the model by the means of prior distributions. Thorlund~et~al.~\cite{thorlund2013modelling} propose to use moderately informative priors instead of uninformative priors when the number of studies is small. This finding was supported by Paul~et~al.~\cite{paul2010bayesian}.  However, assuming inverse gamma priors for the variance parameters and a normal distribution for a transformation of the correlation parameter, it is not clear how to specify the corresponding prior parameters intuitively or how to link them to available expert knowledge.

We propose to use penalised complexity (PC) priors~\cite{2014arXiv1403.4630S}, which we specify separately for the variance and the correlation parameters. 
PC priors follow the principle of shrinking to a so-called base model if not
otherwise indicated by the data. Thinking about the correlation
parameter in most applications we may shrink to zero, which means no
correlation and corresponds to the simplest model. In bivariate
meta-analysis,  we generally may want to shrink to a
negative correlation $\rho_0$ as a base model, which is motivated by a potential threshold variation between studies \cite{glas2003tumor}. Defining the base model $\rho_0$ we can further express our beliefs about deviating from this model in terms of probability
contrasts such as $P(\rho < \rho_0) = \alpha$. Adding one piece of further
information such as a belief about the tail behaviour fully specifies the PC prior. Motivation for the prior distribution for the variances can be linked to a priori range in which we believe sensitivity
or specificity should lay. For example believing that sensitivity lies with
95$\%$ probability in the range $[0.5, 0.95]$ is equivalent to saying that the marginal standard deviation of sensitivity is with $5\%$ probability larger than $3$. This contrast is enough to specify a PC prior for the variance parameter that shrinks to a variance of zero as sensible base model.

We used a simulation study to compare the performance of differently defined PC priors to commonly used prior distributions. The estimates of overall sensitivity and specificity are not sensitive to the choice of different priors for the variance and correlation parameters. Using PC priors leads to a lower MSE and better coverage probabilities for the variance parameters compared to an inverse gamma distribution. When there are only few studies changing the prior distribution for the correlation parameter  will have an effect on the posterior marginals of the correlation. Our findings indicate that no prior is superior in all settings. If the true correlation is in the neighbourhood of the base model more informative PC priors perform better, but their performance decreases if the true correlation is further apart. Considering all scenarios a less informative PC prior performs almost always as good as a normal distribution for the Fisher's z-transformed correlation parameter with mean zero and variance five. While the PC prior performs slightly worse when the true correlation is strongly positive or close to $-1$, it is beneficial for  all other true correlation values. That means the values we regard sensible if not indicated otherwise by the data.

For inference we used integrated nested Laplace approximations, which replaces MCMC sampling by approximating the posterior marginals
directly \cite{paul2010bayesian, rue2009approximate}. This is particularly attractive when many datasets, for example in a simulation study with different scenarios, should be analysed as MCMC would be very time consuming. The recent R-package \texttt{meta4diag} implements Bayesian bivariate meta-analysis based on INLA and offers a
purpose-built user interface to make the full-functionality available
without requiring extensive R knowledge \cite{guo2015}.

Care should be taken when the number of studies involved is really small. The bivariate model is a complex model and data are needed to estimate all present parameters. Agreeing with Takwoingi et al. \cite{takwoingi2015performance} we propose to use a simplified model 
for meta-analyses including few studies, say less than 10. One potential simplification is to specify separate models for sensitivity and specificity which corresponds to
assuming that the correlation parameter for the bivariate random effects is equal to zero. Alternatively, we may fix the correlation to a suitable (negative) value. This feature is also supported in \texttt{meta4diag}.

The prior specification we propose here could also be adapted to trivariate meta-analysis where in addition to sensitivity and specificity also disease prevalence is modelled \cite{chu2009meta}. However, then additional constraints might be necessary for the correlation parameter to ensure that the resulting covariance matrix is positive definite. Another possible extension is to assume that the gold-standard is not perfect \cite{walter1999meta, chu2009random}. However this adds another level to the model and makes it more complex so that parameter estimation gets even more challenging when data are sparse. In future work we would like to analyse sensitivity of PC priors more systematically based on the work of Roos and Held \cite{roos2011sensitivity} and Roos et al. \cite{roos2015sensitivity}.

\section*{Appendix A: Derivation of PC prior for the correlation parameter}
In this appendix, we derive the PC prior for the correlation parameter in the bivariate model. Consider the bivariate random effects model in Equation~(\ref{eq1}), where the bivariate random effect is assumed to be normally distributed with zero-mean and covariance matrix $\mathbf{\Sigma}$. We follow the separation strategy given in Equation \eqref{eq:barnard} and consider only the correlation matrix $\mathbf{R}$ in this section. Let $\mathcal{N}(\mathbf{0}, \mathbf{R}_{b})$ and $\mathcal{N}(\mathbf{0}, \mathbf{R}_{f})$ denote the base and flexible model, respectively
\begin{align}
	\mathbf{R}_{b} = \begin{pmatrix}1 & \rho_0 \\ \rho_0 & 1\end{pmatrix}\quad \quad \quad\quad\quad
 \text{and} \quad \quad \quad\quad\quad
	\mathbf{R}_{f} = \begin{pmatrix}1 & \rho \\ \rho & 1\end{pmatrix},
\label{eq:e2}
\end{align}
where $\rho_0$ is fixed to a (negative) value in which one believes with respect to the application, while $\rho$ is a random variable.

The increased complexity between $\mathcal{N}(\mathbf{0}, \mathbf{R}_f)$
and $\mathcal{N}(\mathbf{0}, \mathbf{R}_b)$ is measured
by the Kullback-Leibler discrepancy (KLD), where the KLD is
\begin{equation*}
\text{KLD}\left( \mathcal{N}(\mathbf{0}, \mathbf{R}_{f}) || \mathcal{N}(\mathbf{0}, \mathbf{R}_{b})  \right) = \frac{1}{2} \left\{ \text{tr}(\mathbf{R}_b^{-1}\mathbf{R}_f) -2-\ln\left(\frac{|\mathbf{R}_f|}{|\mathbf{R}_b|} \right) \right\}.
\end{equation*}
After substituting the equation~(\ref{eq:e2}) into KLD, we get 
\begin{equation*}
\text{KLD}\left( \mathcal{N}(\mathbf{0}, \mathbf{R}_{f}) || \mathcal{N}(\mathbf{0}, \mathbf{R}_{b})  \right)  = \frac{1-\rho_0\rho}{1-\rho_0^2} -1 -\frac{1}{2} \ln (1-\rho^2) + \frac{1}{2} \ln (1-\rho_0^2),
\end{equation*}
and $d(\rho)=\sqrt{ 2\text{KLD}(\rho) }$ is the distance between the base model and flexible model. In order to specify a suitable distribution on the distance, we follow the rules according to \cite{2014arXiv1403.4630S} and assume a constant penalisation, such that the prior satisfies
\begin{equation*}
\frac{\pi_d(d+\delta)}{\pi_d(d)}=r^{\delta}, \quad\quad\quad d,\delta \geqslant 0
\end{equation*}
for some constant decay-rate $0<r<1$. This leads to an exponential prior on the distance scale, $\pi(d) = \lambda\exp(-\lambda d)$, for $r=\exp(-\lambda)$. With a transformation of parameter, we obtain the prior density for $\rho$,
\begin{equation*}
\pi(\rho) = \lambda\exp(-\lambda d(\rho)) \left| \frac{\partial d(\rho)}{\partial \rho} \right|,
\end{equation*}
where 
\begin{equation*}
\frac{\partial d(\rho)}{\partial \rho} = \left(\frac{\rho}{1-\rho^2}-\frac{\rho_0}{1-\rho_0^2}\right)\left/\sqrt{2\text{KLD}}\right..
\end{equation*}
The density implies the identical density behaviour on both sides of $\rho_0$. In order to distribute the density unevenly around of $\rho_0$ but keeping the assumption of a constant rate penalisation, we obtain two one-sided exponential priors $\pi_1(d;\lambda_1)$ and $\pi_2(d;\lambda_2)$ on distance with rate parameters $\lambda_1$ and $\lambda_2$. The functions $\pi_1$ and $\pi_2$ are used to construct the prior for correlation when $\rho\leqslant \rho_0$ and $\rho >\rho_0$, respectively. The parameters $\lambda_1$ and $\lambda_2$ indicate the decay-rate of the deviation from the base model when $\rho\leqslant \rho_0$ and $\rho >\rho_0$, respectively. In order to merge these two exponential priors in the distance scale into one prior distribution $\pi(\rho; \lambda_1,\lambda_2)$ in the correlation scale, we define
\begin{equation*}
\pi(\rho;\lambda_1,\lambda_2) = \left\{
\begin{aligned}
\omega_1\pi_1(d(\rho);\lambda_1)\left| \frac{\partial d(\rho)}{\partial \rho} \right|,\quad \quad \rho\leqslant \rho_0\\
\omega_2\pi_2(d(\rho);\lambda_2)\left| \frac{\partial d(\rho)}{\partial \rho} \right|,\quad \quad \rho> \rho_0
\end{aligned}
\right.
\end{equation*}
where $\omega_1=\int_{-1}^{\rho_0}\pi(\rho;\lambda_1,\lambda_2)$ and $\omega_2 = \int_{\rho_0}^{1}\pi(\rho;\lambda_1,\lambda_2)\in [0, 1]$, are the probability when $\rho\leqslant \rho_0$ and $\rho >\rho_0$, and satisfy $\omega_1+\omega_2=1$ and $\omega_1\lambda_1 = \omega_2\lambda_2$. Thus we have
\begin{equation*}
\omega_1 = \frac{\lambda_2}{\lambda_1+\lambda_2} \quad \quad\quad\quad\quad \text{and} \quad \quad\quad\quad\quad \omega_1 = \frac{\lambda_2}{\lambda_1+\lambda_2}.
\end{equation*}
To specify the hyperparameters $\lambda_1$ and $\lambda_2$, we have to specify a density behaviour on each side of $\rho_0$. The idea behind is to control the prior mass in the tail(s) and(or) the probability that $\rho<\rho_0$. We define 3 strategies to obtained the value of $\lambda_1$ and $\lambda_2$. 

The first strategy is defined under condition $P(\rho\leqslant u_{\text{min}}|-1<u_{\text{min}}<\rho_0)=\alpha_1$ where $u_{\text{min}}$ is a ``left tail event" and $\alpha_1$ is the weight on this event. From the condition, we obtain
\begin{equation*}
\begin{aligned}
&P(\rho\leqslant u_{\text{min}}|-1<u_{\text{min}}<\rho_0)=\alpha_1\\[0.2cm]
\Longrightarrow \quad& \int_{u_{\text{min}}}{\rho_0}\omega_1\pi_1(d(\rho);\lambda_1)\left| \frac{\partial d(\rho)}{\partial \rho} \right| \text{d}\rho = \omega_1-\alpha_1\\[0.2cm]
\Longrightarrow \quad& \int^{d(u_{\text{min}})}_{0}\omega_1\lambda_1\exp(-\lambda_1d)\text{d}d = \omega_1-\alpha_1\\[0.2cm]
\Longrightarrow \quad& \omega_1-\omega_1\exp(\lambda_1d(u_{\text{min}})) = \omega_1-\alpha_1\\[0.2cm]
\Longrightarrow \quad& \lambda_1 = \frac{\log(\omega_1)-\log(\alpha_1)}{d(u_{\text{min}})}.
\end{aligned}
\end{equation*}

The second strategy is defined under condition $P(\rho\geqslant u_{\text{max}}|\rho_0<u_{\text{max}}<1)=\alpha_2$ where $u_{\text{max}}$ is a `right tail event" and $\alpha_2$ is the weight on this event. From the condition, we obtain
\begin{equation*}
\begin{aligned}
&P(\rho\geqslant u_{\text{max}}|\rho_0<u_{\text{max}}<1)=\alpha_2\\[0.2cm]
\Longrightarrow \quad& \int^{u_{\text{max}}}_{\rho_0}\omega_2\pi_2(d(\rho);\lambda_2)\left| \frac{\partial d(\rho)}{\partial \rho} \right| \text{d}\rho = \omega_2-\alpha_2\\[0.2cm]
\Longrightarrow \quad& \int^{d(u_{\text{max}})}_{0}\omega_2\lambda_2\exp(-\lambda_2d)\text{d}d = \omega_2-\alpha_2\\[0.2cm]
\Longrightarrow \quad& \omega_2-\omega_2\exp(\lambda_2d(u_{\text{max}})) = \omega_2-\alpha_2\\[0.2cm]
\Longrightarrow \quad& \lambda_2 = \frac{\log(1-\omega_1)-\log(\alpha_2)}{d(u_{\text{max}})}.
\end{aligned}
\end{equation*}

The third strategy is defined under conditions $P(\rho\leqslant u_{\text{min}}|-1<u_{\text{min}}<\rho_0)=\alpha_1$ and $P(\rho\geqslant u_{\text{max}}|\rho_0<u_{\text{max}}<1)=\alpha_2$ without giving any information about $\omega$'s. In order to obtain the value for $\lambda_1$ and $\lambda_2$, we solve the following equations numerically
\begin{equation*}
\left\{
\begin{aligned}
\quad&\lambda_1 = \frac{\log(\omega_1)-\log(\alpha_1)}{d(u_{\text{min}})}\\[0.2cm]
\quad&\lambda_2 = \frac{\log(1-\omega_1)-\log(\alpha_2)}{d(u_{\text{max}})}\\[0.2cm]
\quad&\omega_1\lambda_1 = (1-\omega_1)\lambda_1.
\end{aligned}
\right.
\end{equation*}

\section*{Appendix B: R-code}
Before using the package \texttt{meta4diag} \cite{guo2015}, the package INLA needs to be loaded in R:
\begin{verbatim}
R> library(INLA)
R> library(meta4diag)
\end{verbatim}
The telomerase data set of Glas et al. \cite{glas2003tumor} is included in the \texttt{meta4diag} package and can be loaded with
\begin{verbatim}
R> data(Telomerase)
\end{verbatim}
The first six lines are as follows
\begin{verbatim}
R> head(Telomerase)
  References No.Study TP FP FN  TN 
1        Ito       59 25  1  8  25 
2      Rahat       35 17  3  4  11 
3    Kavaler      151 88 16 16  31 
4    Yoshida      109 16  3 10  80 
5  Ramakumar      195 40  1 17 137
6    Landman       77 38  6  9  24 
\end{verbatim}

\begin{figure}[h!]
\centering
\includegraphics[width=0.8\textwidth]{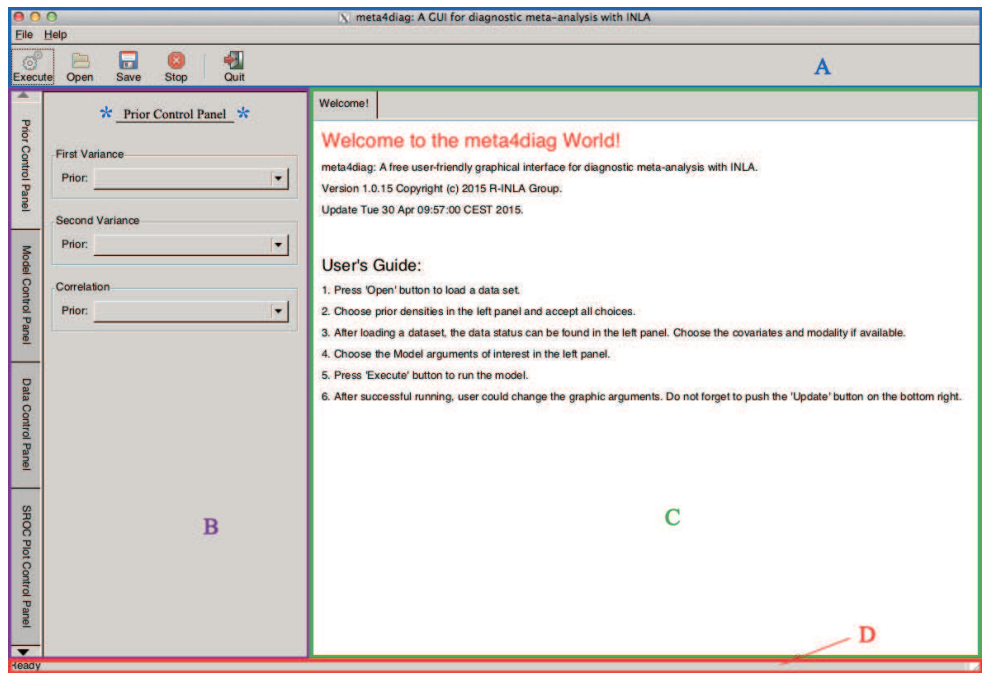}
\caption{GUI main window of \texttt{meta4diag} after start up. (A) Menu bar and toolbar, (B) tool
panel, (C) view area, showing different pages (welcome message, data set, summary results, graphics) and (D) status bar.}
\label{guo10}
\end{figure}

To analyse the telomerase data, we can either use the graphical user interface, see Figure~\ref{guo10} or use the R terminal. Here we explain all R commands. The main function in the package is \texttt{meta4diag()} which we use as follows
\begin{verbatim}
R> res = meta4diag(data = Telomerase,
+    var.prior = "PC", var2.prior = "PC", cor.prior = "PC",
+    var.par = c(3, 0.05), var2.par = c(3, 0.05),
+    cor.par = c(1, -0.2, 0.4, -0.95, 0.05, NA, NA))
\end{verbatim}
The first argument is the telomerase dataset. The arguments ``var.prior", ``var.par", ``var2.prior", ``var2.par", ``cor.prior" and ``cor.par" are used to specify the prior distributions for all hyperparameters. Here, the arguments ``var.prior'', ``var2.prior'' and ``cor.prior'' are strings specifying the name of prior density for the first variance component, the second variance component and the correlation, respectively. Possible choices are for example ``InvGamma" or ``PC" for the inverse gamma distribution for the variance parameter or the PC prior. Here, we use PC priors. The argument ``var.par'' is a numerical vector specifying the parameters of the prior for the first variance and is defined as var.par~=~c($u, \alpha$) when ``var.prior'' is ``PC''. The parameters of the prior for the second variance are analogously specified. The argument ``cor.par'' is a numerical vector to specify the parameters of the prior for the correlation parameter and defined as cor.par~=c(strategy, $\rho_0$, $\omega_1$, $u_{\text{min}}$, $\alpha_1$, $u_{\text{max}}$, $\alpha_2$) when ``cor.prior'' is ``PC''. Here, we implement the same prior settings as in Section 5, i.e., strategy 1 described in Section 3.

A summary of the fitted model with estimates $\mu$, $\nu$ as well as logit$^{-1}(\mu)$ and logit$^{-1}(\nu)$ and hyperparameters $\sigma^2_{\phi}$, $\sigma^2_{\psi}$ and $\rho$ can be obtained with:
\begin{verbatim}
R> summary(res)
Time used: 
 Pre-processing    Running inla Post-processing 
     0.71966100      0.44033289      0.09974599 
          Total 
     1.25973988 

Fixed effects: 
    mean    sd 0.025quant 0.5quant 0.975quant
mu 1.192 0.197      0.806    1.189      1.592
nu 2.289 0.638      1.086    2.264      3.639

Model hyperpar: 
           mean    sd 0.025quant 0.5quant 0.975quant
var_phi   0.237 0.178      0.046    0.188      0.709
var_psi   3.491 1.972      1.133    2.996      8.683
cor      -0.791 0.217     -0.995   -0.865     -0.197

-------------------
          mean    sd 0.025quant 0.5quant 0.975quant
mean(Se) 0.766 0.029      0.706    0.767      0.820
mean(Sp) 0.898 0.047      0.785    0.906      0.966

-------------------
Correlation between mu and nu is -0.4702.
Marginal log-likelihood: -65.4381
Variable names for marginal plotting: 
      mu, nu, var1, var2, rho
\end{verbatim}
The SROC curve shown in Figure~\ref{guo9} is obtained using 
\begin{verbatim}
R> ROC(res)
\end{verbatim}
More details on how to extract parameter estimates and generate other graphics of interest are given in Guo and Riebler \cite{guo2015}.

\bibliographystyle{plain}
\bibliography{metaPC}

\newpage
\begin{landscape}
\begin{table}[h!]
\caption{Three different strategies to specify a PC prior for the correlation parameter $\rho$. Shown are the corresponding probability contrasts that need to be provided.}
\label{table:strategies}
\centering
\begin{tabular}{@{\extracolsep{4pt}}cccc@{}}
\toprule
\multirow{2}{*}{\textbf{Strategy}}  &  \multicolumn{3}{c}{\textbf{Condition and parameters}}    \\
\cline{2-4}
 & $P(\rho<u_{\text{min}} | -1<u_{\text{min}}<\rho_0)=\alpha_{1}$   & $P(\rho>u_{\text{max}}| \rho_0<u_{\text{max}}<1)=\alpha_{2}$ & $P(\rho \leqslant \rho_{0})=\omega_1$  \\
\midrule
1 &  \checkmark  &  & \checkmark \\
2 &   &  \checkmark & \checkmark  \\
3 & \checkmark  &  \checkmark &    \\
\bottomrule
\end{tabular}
\end{table}

\end{landscape}

\newpage
\begin{table}
\caption{The different scenarios used in the simulation study. Each subset of nine scenarios corresponds to the correlation values $\rho=-0.95, -0.8, -0.6, -0.4, -0.2, 0, 0.2, 0.4, 0.6$.}
\label{table:table1}
\centering
\begin{tabular}{lccccc}
\toprule
Scenario & logit$^{-1}(\mu)$ & logit$^{-1}(\nu)$ & $\sigma_{\phi}^2$ & $\sigma_{\psi}^2$ & $I$ \\
\midrule
1-9          &    0.8      &     0.7     &    1        &     1      &   10       \\ 
10-18        &    0.8      &     0.7     &    1        &     1      &   25       \\ 
19-27        &    0.8      &     0.7     &    1        &     1      &   50      \\ 
                 &               &               &              &             &              \\
28-36        &    0.9       &     0.9    &    1        &     1      &   10       \\ 
37-45        &    0.9       &     0.9    &    1        &     1      &   25       \\
46-54        &    0.9       &     0.9    &    1        &     1      &   50       \\  
                  &                &              &             &             &               \\
55-63         &    0.95     &     0.3    &    1       &     1      &   10       \\ 
64-72         &    0.95     &     0.3    &    1       &     1      &   25       \\ 
73-81         &    0.95     &     0.3    &    1       &     1      &   50       \\ 

\bottomrule
\end{tabular}
\end{table}

\newpage
\begin{table}[h!]
\caption{Meta-analysis in which the telomerase marker is used to diagnose bladder cancer.}
\label{table:realdata}
\centering
\begin{tabular}{lcccc}
\toprule
   Study           &      TP      & FN     &     TN     &    FP  \\
 \hline
Ito et al. & 25 & 8 & 25 & 1  \\
Rahat et al. & 17 & 4  &11 & 3  \\
Kavaler et al. & 88 & 16 & 31 &  16  \\
Yoshida et al. & 16 & 10 & 80 & 3  \\
Ramakumar et al. & 40 & 17 & 137 & 1   \\
Landman et al. & 38 & 9 & 24 & 6  \\
Kinoshita et al. & 23 & 19 & 12 & 0  \\
Gelmini et al. & 27 & 6 & 18 & 2  \\
Cheng et al. & 14 & 3 & 29 & 3  \\
Cassel et al. & 37 & 7 & 7 & 22  \\
\bottomrule
\end{tabular}
\end{table}

\end{document}